\documentclass[sigconf]{acmart}
\def\BibTeX{{\rm B\kern-.05em{\sc i\kern-.025em b}\kern-.08emT\kern-.1667em\lower.7ex\hbox{E}\kern-.125emX}}
    
\copyrightyear{}
\acmYear{}
\acmConference[]{}{}{}
\acmBooktitle{}
\acmPrice{}
\acmDOI{}
\acmISBN{}

\renewcommand\footnotetextcopyrightpermission[1]{}
\begin{document}

%
% The "title" command has an optional parameter, allowing the author to define a "short title" to be used in page headers.
\title[CANet]{CANet: An Unsupervised Intrusion Detection System for High Dimensional CAN Bus Data} %: A Novel Neural Network Architecture

%
% The "author" command and its associated commands are used to define the authors and their affiliations.
% Of note is the shared affiliation of the first two authors, and the "authornote" and "authornotemark" commands
% used to denote shared contribution to the research.
%\author{Markus Hanselmann}
%\authornote{Contributed equally.}
%\email{markus.hanselmann@etas.com}
%\orcid{}
%\email{webmaster@marysville-ohio.com}
%\affiliation{%
%  \institution{Institute for Clarity in Documentation}
%  \streetaddress{P.O. Box 1212}

\author{Markus Hanselmann}
\authornote{Contributed equally.}
\affiliation{%
  \institution{ETAS GmbH, Bosch Group}
  \streetaddress{}
  \city{Stuttgart}
  \country{Germany}}
\email{markus.hanselmann@etas.com}

\author{Thilo Strauss}
\authornotemark[1]
\affiliation{%
	\institution{ETAS GmbH, Bosch Group}
	\streetaddress{}
	\city{Stuttgart}
	\country{Germany}}
\email{thilo.strauss@etas.com}

\author{Katharina Dormann}
\affiliation{%
 \institution{Robert Bosch GmbH}
 \streetaddress{}
 \city{Ludwigsburg}
 \country{Germany}}
\email{katharina.dormann@de.bosch.com}

\author{Holger Ulmer}
\affiliation{%
	\institution{ETAS GmbH, Bosch Group}
	\streetaddress{}
	\city{Stuttgart}
	\country{Germany}}
\email{holger.ulmer@etas.com}

%
% By default, the full list of authors will be used in the page headers. Often, this list is too long, and will overlap
% other information printed in the page headers. This command allows the author to define a more concise list
% of authors' names for this purpose.
\renewcommand{\shortauthors}{Hanselmann and Strauss, et al.}

%
% The abstract is a short summary of the work to be presented in the article.
\begin{abstract}
We propose a novel neural network architecture for detecting intrusions on the CAN bus. The Controller Area Network (CAN) is the standard communication method between the Electronic Control Units (ECUs) of automobiles. However, CAN lacks security mechanisms and it has recently been shown that it can be attacked remotely. Hence, it is desirable to monitor CAN traffic to detect intrusions. In order to detect both, known and unknown intrusion scenarios, we consider a novel unsupervised learning approach which we call CANet. To our knowledge, this is the first deep learning based intrusion detection system (IDS) that takes individual CAN messages with different IDs and evaluates them in the moment they are received. This is a significant advancement because messages with different IDs are typically sent at different times and with different frequencies. Our method is evaluated on real and synthetic CAN data. For reproducibility of the method, our synthetic data is publicly available. A comparison with previous machine learning based methods shows that CANet outperforms them by a significant margin. 
\end{abstract}

\maketitle

\section{Introduction}

\begin{table*}[t]
	\centering
	\begin{tabular}{cc|cccccc|ccc|ccc|c}
		\toprule
		\multicolumn{15}{c}{CAN Bus Data after Preprocessing}\\
		%\toprule
		%\multicolumn{2}{c}{Test Data}   & \multicolumn{2}{c}{No Attack}  & 
		%\multicolumn{2}{c}{Grad. 1}&Grad. 2\\
		%\midrule
		% &Method  &Single& Ensemble  &Single& Ensemble  & Ensemble\\   \midrule
		\midrule
		%Time Step  & ID & \multicolumn{13}{c}{Signal Values}  \\   \midrule
		Time Stamp  & ID & \multicolumn{6}{c}{Signals of $A$} & \multicolumn{3}{c}{Signals of $B$} & \multicolumn{3}{c}{Signals of $C$} & Signals of $D$ \\   \midrule 
		1.045  &B & - & - & - & - & - & - & 54.71 & 0 & 7.24 & - & - & - & -\\
		3.102 &D & - & - & - & - & - & - & - & - & - & - & - & - & 31.47\\
		4.978  &A & 12 & 44.15 & 38.02 & 2 & 0 & 1 & - & - & - & - & - & - & -\\
		7.014  &C & - & - & - & - & - & - & - & - & - & 17.79 & 7 & 2 & -\\
		8.993  &B & - & - & - & - & - & - & 55.02 & 1 & 7.21 & - & - & - & -\\
		9.750  &A & 13 & 44.01 & 39.67 & 1 & 0 & 2 & - & - & - & - & - & - & -\\
		\vdots  &\vdots & \vdots  & \vdots & \vdots  & \vdots & \vdots & \vdots & \vdots & \vdots & \vdots & \vdots & \vdots & \vdots & \vdots\\
		
		\bottomrule \\
	\end{tabular}
	\caption{Schematic representation of CAN bus data after preprocessing its bytes to signals. Note, that at each time stamp only the signal values of a single ID are transmitted. Different IDs may contain a different number of signals, e.g. ID $A$ consists of six signals whereas ID $D$ has one signal. The time stamp is given in milliseconds. In CANet, the time is discretized. } 
	\label{CANdata}
\end{table*}

Automobiles are getting more and more connected by technologies such as Bluetooth, Wifi or smart phone plug-ins. While this simplifies the driver's life, it simultaneously opens new paths for potential remote attacks on the Electronic Control Units (ECUs) of cars. Hijacking an ECU can allow attackers to place messages on the vehicle-internal communication network and, thus, e.g. to invoke sudden breaking or turning off the engine which can, potentially, cause traffic accidents~\cite{checkoway2011comprehensive, miller2015remote}. This may have even more disastrous outcomes in autonomous vehicles. Hence, detecting the attempt of attacks in car networks is in the interest of traffic safety. %For demonstrated example attacks on cars see~\cite{checkoway2011comprehensive, miller2015remote}. 

In this paper, we focus on the CAN bus as it is the most common vehicle bus standard. Typically, CAN messages are used to transmit signals between ECUs. For example, an ECU can send the information about objects on the road so that the break assist can react accordingly. An extensive overview about previous work on CAN intrusion detection systems can be found in ~\cite{tomlinson2018towards}. A strong focus lies on rule based and statistical methods to detect known attack scenarios. While many types of intrusions can be detected efficiently by these approaches, the configuration of such an IDS is time-consuming, requires domain expertise, and it is unlikely that unknown attack scenarios can be detected. Moreover, it is challenging to generate rules that capture the underlying behavior of signals or physical dependencies between them. 

With the advances in deep learning in the recent years~\cite{krizhevsky2012imagenet, lecun2015deep, hinton2012deep}, new tools are becoming available that have the potential of detecting unknown attacks. Prior work on intrusion detection with neural networks on single CAN signals can be found in the literature \cite{taylor2016anomaly, WeberWolfSax2018_1000084311, weber2018embedded}. However, to the best of our knowledge, there is no neural network architecture that can handle the CAN bus data structure in the signal space. On the CAN bus, at every point in time at most one message is transmitted. As a result the CAN traffic consists of consecutive messages with different IDs. These messages contain different kinds of signals  (see Table \ref{CANdata}). The data structure makes it difficult to feed the data of the CAN bus directly into any kind of standard neural network. 

The contributions of this manuscript are the following: We introduce CANet, a novel neural network architecture tailored to work on the signal space of CAN data and we show that it outperforms baseline methods by a significant margin. For each message ID we introduce one separate long short-term memory (LSTM) \cite{hochreiter1997long}  subnetwork with input dimension equal to the number of signals from the message of that particular ID. The output of the LSTM subnetworks is concatenated to a single latent vector that encodes the current state of  the entire CAN traffic. This vector is followed by several fully connected layers in an autoencoder setting to finally reconstruct the payload of the input message. We describe how this architecture can be trained in an unsupervised manner and evaluated in such a way that a variety of unknown attack types can be detected while identifying normal data correctly. We show that our method is especially strong in finding certain manipulations of signals that are difficult to detect by classical approaches (e.g. the continuous change of a signal towards a desired value). Hereby, we exploit the fact that by processing the joint latent vector the network is able to learn the functional dependencies of the signals under consideration. 

We point out that anomaly detection in the signal space of CAN bus data may have other applications beyond intrusion detection e.g. early detection of technical failures.

This document is organized in the following way: In Section~\ref{Background}, the background of the CAN bus and the relevant literature on CAN IDS is briefly covered. In Section~\ref{netArch}, the proposed network architecture and its training process is described.  In Section~\ref{exps}, the method is evaluated on real and synthetic CAN data and compared to related work. Finally, in Section~\ref{conc}, we present our conclusions and future work.

\section{Relevant Background}\label{Background}
\subsection{Terminology}
The Controller Area Network (CAN) is a vehicle bus standard designed to allow automotive Electronic Control Units (ECUs) to communicate with each other. 

A CAN message is characterized by a time stamp, an ID and typically an 8-byte payload field. The ID represents the type of the current message and the payload field is used to communicate current values of vehicle signals. Each ID is associated with a set of signals and the payload of messages carrying this ID provides their current values. In general, messages with different IDs contain a different set of signals. The encoding of signal values ranges from single bits to several bytes of the payload. A so-called CAN matrix provides information for each ID, specifying which payload bits encode which signal. In this paper, we assume that the decoding of raw payload bits into signal values is already performed (see Table~\ref{CANdata}). 

\subsection{Objective}
The objective is to detect intrusions into the CAN bus communication. We assume that the attacker has already gained access to the CAN bus, e.g. by hijacking one of the connected ECUs. This follows the path of demonstrated security relevant attacks \cite{miller2015remote}. We further assume that an attacker now tries to influence the vehicle behavior by manipulating messages. 

Our goal is to detect signals deviating from their ``normal'' behavior or signals breaking out of physical relationships. These physical relationships are typically complex, potentially unknown and hard to derive by rule based intrusion detection systems for CAN.

\subsection{Related Research}\label{related}

In this section, we discuss previous approaches for intrusion detection on CAN bus data. In \cite{taylor2016anomaly}, a LSTM network structure for predicting the next payload of a single ID is proposed. Other deep learning based results include \cite{ WeberWolfSax2018_1000084311}, where an autoencoder like network architecture is used on a sliding window over the appearances of a  specific signal.  In \cite{weber2018embedded}, a lightweight on-line detector of anomalies (LODA) is proposed. While these methods show promising results, they all  build a model for a single time series containing the values from one signal or signals of one ID. We extend this by enabling the network to work with all signals of multiple CAN IDs simultaneously. This gives CANet the advantage that it can detect intrusions by inferring functional or physical dependencies that appear between signals on the CAN bus.

Non neural network based methods for anomaly detection on the CAN bus payload include 
signature based methods \cite{studnia2014language},
finger printing \cite{cho2016fingerprinting},
clustering methods \cite{tomlinson2018using},
fuzzy logic \cite{martinelli2017car},
Hidden-Markov-Model based methods \cite{narayanan2016obd_securealert},
and entropy based methods \cite{muter2011entropy, marchetti2016evaluation}. A comprehensive review of the strengths and weaknesses of these and other non-payload based methods can be found in \cite{tomlinson2018towards}.

\section{CANet}\label{netArch}

\begin{figure*}
	\begin{center}
		\includegraphics[width=\textwidth]{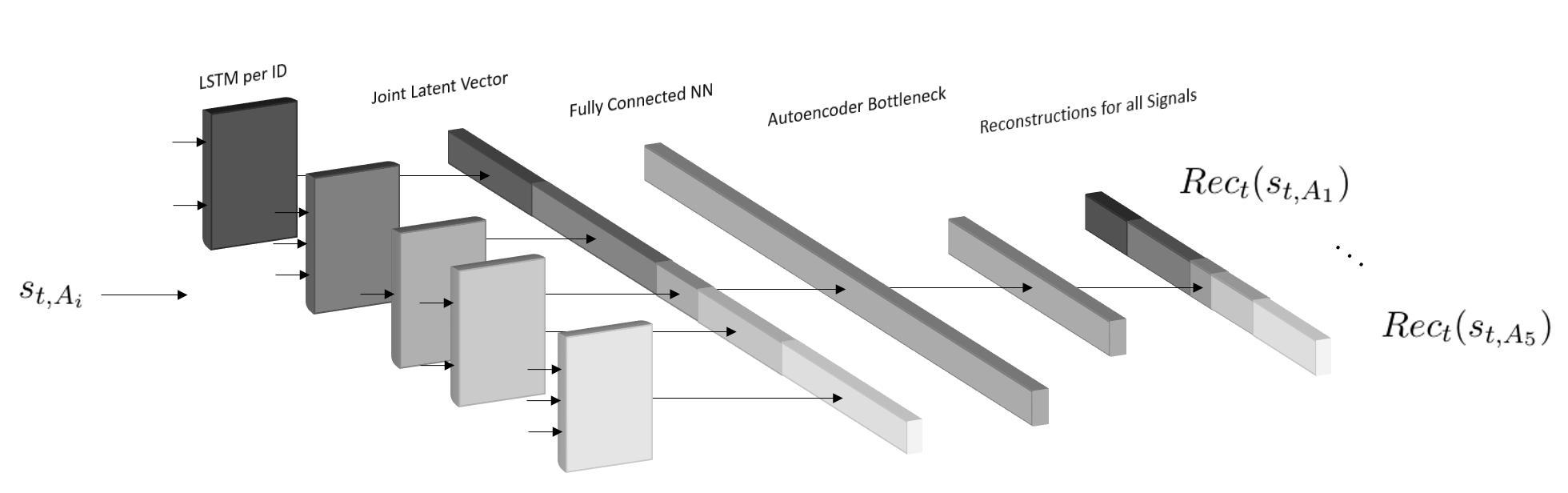}
	\end{center}
	\caption{Schematic representation of the CANet architecture for intrusion detection on CAN bus data. Each ID has its own LSTM input model. When the payload $s_{t,A_i}$ of an ID is fed into its input model, only the corresponding memory in the joint latent vector is updated. The entire latent vector is used to reconstruct the signals of all IDs. The deviation between the true input payload and the corresponding reconstruction is used for anomaly detection.}
	\label{NN_arch}
\end{figure*}

The basic idea of CANet is to handle the challenging structure of CAN data by introducing several independent recurrent neural networks on the input side. This enables the network to learn temporal dependencies in the signals. In detail, we introduce a separate LSTM for each ID that gets the signal values associated to this ID as inputs and that stores the current state of the processed data. The state of the whole CAN traffic is then represented by a joint latent vector that is realized as concatenation of the current states of all input networks. This vector does not contain any information about what ID has been processed last. To enable an unsupervised learning setting, the joint latent vector is fed into a subnetwork of consecutive linear layers in an autoencoder setting. That is, at each time step the task of this subnetwork is to reconstruct the signal values of each possible input message solely based on the current joint latent vector. The deviation between the signal values of the true input message and its reconstruction is then used as a measure for the normality of the input message. Since the network is solely trained on normal data, it is expected that this reconstruction error is small on normal data and large on anomalous data.

\subsection{Network Architecture}

In order to describe the network architecture efficiently, we first establish some notations. Let ${\mathbf{A} = \{A_{1}, \dots , A_{K}\}}$ be the ordered set of all $K \in \mathbb{N}$ considered IDs. For each ID $A\in\mathbf{A}$ we take $n_{A}$ corresponding signals into account that are encoded in the payload. We denote the total number of signals by $N$. Whenever an ID $A\in\mathbf{A}$ is observed at time step $t$, we denote the vector containing the corresponding signal values by $s_{t, A}\in\mathbb{R}^{n_{A}}$. Note that we hereby discretize the time.

A visualization of the architecture can be found in Figure \ref{NN_arch} and the corresponding layer specifications in Table \ref{net_Table}. The network architecture consists of an input LSTM subnetwork for every ID ${A\in\mathbf{A}}$. The $i$-th LSTM subnetwork is associated with  ID $A_i$, has  $n_{A_i}$ inputs, and a hidden dimension of size ${n_{A_i}\cdot h_{scale}}$. Here, ${h_{scale}\in \mathbb{N}}$ represents the computational power of CANet. In the evaluation part, the performance with different $h_{scale}$ values is compared (see section \ref{eval_sec}). Whenever a new payload of an ID ${A\in\mathbf{A}}$ is fed through the corresponding LSTM subnetwork, its output of size ${n_{A}\cdot h_{scale}}$ is used to update the corresponding memory in the joint latent vector. Hence, the joint latent vector has length ${N \cdot h_{scale}}$. It represents and stores the current state of the CAN traffic. The joint latent vector is followed by a set of fully connected layers where the penultimate layer has strictly less neurons than the output layer, which has $N$ neurons. The task of the output layer is to reconstruct all potential current input signals from all IDs.

\begin{table}[t]
	\centering
	\begin{tabular}{lll}
		\toprule
		\multicolumn{2}{c}{CANet Architecture}\\
		%\toprule
		%\multicolumn{2}{c}{Test Data}   & \multicolumn{2}{c}{No Attack}  & 
		%\multicolumn{2}{c}{Grad. 1}&Grad. 2\\
		%\midrule
		% &Method  &Single& Ensemble  &Single& Ensemble  & Ensemble\\   \midrule
		\midrule
		Layer Type & Size \\ \midrule
		LSTM per ID &  Number of Signals of ID $\times$ $h_{scale}$ \\
		Joint Latent Vector & $N \times h_{scale}$ \\
		Fully Connected (ELU) & $N \times h_{scale}/2$ \\
		Fully Connected (ELU) & $N-1$ \\
		Fully Connected (ELU) & $N$  \\
		\bottomrule \\
	\end{tabular}
	\caption{CANet architecture for detecting intrusions on the CAN bus. A total number of $N$ signals is considered. The parameter  $h_{scale}\in\mathbb{N}$ determines the computational power of the neural network and is specified in the evaluation part.} 
	\label{net_Table}
\end{table}

During training, at each time step $t\in\mathbb{N}$ the payload of an ID, $A_i$ say, is fed through its corresponding LSTM input model. It then is used to update the joint latent vector in order to reconstruct the payload at time step $t$ for all $A\in\mathbf{A}$. Formally, the reconstruction $R_t$ of the payload for the time step $t$ is denoted by 
\begin{equation*}
R_{t}(s_{t,A_i}) = ( Rec_{t}(s_{t,A_1}), \dots ,   Rec_{t}(s_{t,A_K}) ),
\end{equation*}
where $Rec_{t}(s_{t,A_j})$ denotes the reconstruction of the payload associated to ID $A_j$.

We then compare the true signal values from the payload of the current ID $A_i$ with their reconstructions $Rec_{t}(s_{t,A_i})$. We use the quadratic error loss function, given by 
\begin{equation}\label{loss}
loss(s_{t, A_i}) = ||Rec_{t}(s_{t,A_i})-s_{t, A_i}||_{\ell_2}^2.
\end{equation}
In detail, when computing the back propagation with respect to this loss function, only the gradients of the weights for the LSTM subnetwork of ID $A_i$ and the weights that connect the joint latent vector with the output $Rec_{t}(s_{t, A_i})$ must be computed.

Since the temporal dependencies are stored for each ID separately in the corresponding LSTM subnetwork, the training process has the advantage that the model as a whole is not sensitive to the exact order of consecutive message IDs. In fact, in real CAN data, there is some variability in the order of IDs, even within the same data set.

\subsection{Anomaly Score}\label{anomalyScore}

The quadratic error between the signal and its reconstruction as defined in Equation~\ref{loss} can be used to predict whether or not a signal at time step $t\in \mathbb{N}$ is anomalous. This prediction can  be made by testing if the error is above a fixed threshold. Due to the fact that we consider an unsupervised learning problem, the threshold must be chosen solely based on normal data. It is computed for each signal separately and is given by the $99.99\%$ percentile of all corresponding quadratic errors on a validation data set. During evaluation, an individual anomaly indicator is stored for every signal. Every time an ID is processed by the model, the anomaly indicator of the respective signals is updated and set to 1 if the reconstruction error exceeds the corresponding $99.99\%$ percentile and set to 0 otherwise. The global anomaly score at a time step $t$ is set to 1 if and only if at least one of the stored signal anomaly indicators is 1 and set to 0 otherwise. 

Note that this anomaly score is only feasible if the number of signals does not get too large because it suffers from similar effects as multiple testing problems. That is, if the number of signals becomes too large, the probability that a normal data point is identified as such (true negative) decreases. %That is, for $n$ dependent events, as it is the case in our decision method, the probability that a normal data point is detected as a normal data point can be bounded below with Boole's inequality by $1 - n (1 - 0.9999)$. This is not problematic for small numbers of $id$'s for example with $n=20$ where lower bound bounded would be $0.998$. However, it might produce considerable to many normal data points detected as abnormal when $n$ is large. For example with $N=1000$ the probability for detecting a normal point as normal is only bounded below by $0.9$. 

While this anomaly score is good for demonstrating the capabilities of our method, it may need to be refined for in-vehicle usage, where a 100\% true negative rate is required because the cost of each false positive is relatively high. That is, depending on the response mechanism this might have consequences such as recommending the driver to stop or see a mechanic due to an attack.

\section{Experiments}\label{exps}

In this section, we evaluate our method on both, real and synthetic CAN data. 

The real data was collected on a test vehicle. In our experiments, 13 IDs with a total number of 20 signals are taken into consideration. The signals are chosen in such a way that they contain physical values and that, for each signal, there is at least one other signal with a  functional dependency to it. We divide about 13 hours of recorded data into 12.5 hours of training and 0.5 hours of test data. We only consider data representing the normal driving mode. Hence, we exclude e.g. starting and turning off the engine. All payloads are preprocessed into their signal value space (see Table \ref{CANdata}). 

In the case of the synthetic data, we consider a data set consisting of 10 different message IDs, each with different amounts of signals per ID and different noisy time frequencies. The total amount of signals is 20. The data is created in such a way that it is similar, based on our experience, to real CAN traffic. The data contains physical values, counters and signals that are dependent on one or multiple other signals. We use a training data set of about 16.5 hours and a test data set of about 7.5 hours of CAN traffic. The data set is available at \url{https://github.com/etas/SynCAN}.

\subsection{Simulated Attacks}

In both, the real and the synthetic data set, the test data is divided into six subsets of equal time length. We use one subset to evaluate our model on normal data. The other five test data sets are used to evaluate our model on the following attack types:
\begin{enumerate}
	\item \textit{Plateau attack}: A single signal is overwritten to a constant value over a period of time, i.e. a jump or freezing the signal. We only consider jumps in the typical signal range. Higher jumps represent a clear attack. For example, a car cannot speed up from $20$ km/h to $100$ km/h within 10 ms. Such attacks might be detected just by considering the respective signal. Smaller jumps or freezing might only be detected if we consider a set of signals with some kind of correlation between them.
	\item \textit{Continuous change attack}: A signal is overwritten so that it slowly drifts away from its true value. This assumes that the attacker wishes to set a signal to a concrete value while trying to fool the IDS with realistic small changes in the signal.
	\item \textit{Playback attack}: A signal value is overwritten over a period of time with a recorded time series of values of that signal. The attacker hopes to trick the IDS by sending completely real looking signal values of a different traffic situation. 
	\item \textit{Suppress attack}: The attacker prevents an ECU from sending messages, for example, by turning it off. This kind of attack means that messages of some particular ID do not appear in the CAN traffic for some period of time. 
	\item \textit{Flooding attack}: The attacker sends messages of a particular ID with high frequency to the CAN bus. This attack is easier to perform in praxis then the aforementioned ones, since the attacker does not need to control an ECU. It only requires to send additional messages to the CAN bus in order to ``overwrite'' the real message values. 
\end{enumerate}
The length of a typical attack interval,  in both the real and the synthetic data set, is between 2-4 seconds. In each synthetic test data set are about $100$ and in each real test data set about $10$ attack intervals of the corresponding type. The suppress and flooding attack can be relatively easily detected with a rule based method by analyzing the frequencies of the signals. However, the plateau, continuous change and playback attack are rather difficult to be detected with a rule based system. Furthermore, for all three attacks it is often not sufficient to only consider each signal separately. For example, in the playback attack it is mostly impossible to detect the majority of an attack interval if no access to some correlated signals is given.   

\subsection{Network Training Details}

In this section, we present the training details of our method in order to make the results reproducible. All code for training and evaluation is written with pyTorch \cite{paszke2017automatic}. We use the network architecture described in Table \ref{net_Table} for different $h_{scale}$ values. The optimizer of choice is Adam \cite{kingma2014adam} with a initial learning rate of $0.01$. The data is signal wise 0-1 scaled. We train the network for 1000 epochs with batch size 25. Every element in a batch is a series of 5000 consecutive messages at random starting position in the training data. At the beginning of each epoch the hidden and cell state vector of all LSTM models are initialized with zero. During a single epoch, a back-propagation is performed every $250$ iterations (time steps) in order to update the network weights. For a more robust training, the loss function is multiplied by a fixed scalar for different IDs. That is, the scalar is linearly smaller the more frequent its corresponding ID appears in the training data set. This is to ensure that IDs that appear more often do not get more weight than less frequent IDs during training. 

The training is performed on the training data set after removing a small portion of the data that is used to compute the thresholds for the anomaly score.

\subsection{Comparison with Related Research}
CANet is the first approach capable of handling the data structure of CAN bus data with multiple CAN IDs simultaneously within a single neural network model. Therefore, a one to one comparison with an existing method is not possible. In order to compare CANet with a baseline, we adapted the following methods:
\begin{enumerate}
	\item \textit{Predictive Baseline}: In \cite{taylor2016anomaly}, the basic idea is to learn a separate model for each ID. At each time step, the model predicts the payload of the next occurrence of its associated ID. The network directly processes the bit representation of the payload. As preprocessing step, the raw data is fed to a subnetwork of linear layers. The output is then processed by a combination of LSTM and linear layers that perform the prediction. The difference between the true value and the prediction is used for the anomaly score. We train one predictive model per ID. Since we have access to the signal representation of the payload, we omit the preprocessing subnetwork and feed the signal values directly into the LSTM layers that are followed by a set of linear layers.
	\item \textit{Autoencoder Baseline}: In \cite{ WeberWolfSax2018_1000084311}, an autoencoder model for a single signal is used. That is, at time step $t$ the network has the task to reconstruct the input vector that consists of the signal values on a sliding window at the time steps $(t-7, \dots, t)$. We use their network architecture to obtain one model for each signal.
\end{enumerate}
Following our approach in section \ref{anomalyScore}, we use the $99.99\%$ percentile of the quadratic errors on a validation set (signal wise) to combine it to a final anomaly score.

\begin{table*}[t]
	\centering
	\begin{tabular}{lllllllll}
		\toprule
		\multicolumn{8}{c}{Evaluation Table}\\
		\midrule
		\midrule
		\multicolumn{8}{c}{Synthetic Data}\\
		%\toprule
		%\multicolumn{2}{c}{Test Data}   & \multicolumn{2}{c}{No Attack}  & 
		%\multicolumn{2}{c}{Grad. 1}&Grad. 2\\
		%\midrule
		% &Method  &Single& Ensemble  &Single& Ensemble  & Ensemble\\   \midrule
		\midrule
		\multicolumn{2}{c}{Model Specification}& Accuracy& \multicolumn{5}{c}{True Positive Rate / True Negative Rate}\\
		\midrule
		Method  & $h_{scale}$ & No Attack  & Plateau & Continuous  & Playback & Suppress & Flooding \\    \midrule
		CANet  &5  & 0.991  & 0.896 / 0.980  & 0.740 / 0.994   &  0.896 / 0.997  &  0.496 / 0.996 &  0.900 / \textbf{0.997} \\
		CANet  &10 & 0.990   &  \textbf{0.955} / 0.975  & 0.765 / 0.994  &  0.905 / 0.996  & \textbf{0.613} / 0.996  & \textbf{0.901} / 0.996  \\ 
		CANet  &20 &  0.992 & 0.885 / 0.993  & \textbf{0.771} / 0.996  &\textbf{ 0.906} / 0.997 & 0.581 / 0.995  & 0.884 / \textbf{0.997} \\
		CANet  &30 &0.995  & 0.707 / \textbf{0.996}  & 0.446 / \textbf{0.997} & 0.630 / \textbf{0.998}  &0.184 / \textbf{0.997}  & 0.625 / \textbf{0.997}  \\
		\midrule 
		Predictive &- &\textbf{0.996}  & 0.330 / 0.974  & 0.015 / 0.994 & 0.020 / 0.996  & 0.003 / 0.993  & 0.644 / 0.994  \\
		Autoencoder &- & 0.983 &  0.355 / 0.927  &  0.016 / 0.975  &  0.029 / 0.995    &  0.001 / 0.993   &  0.688 / 0.995  \\
		\midrule
		\multicolumn{8}{c}{Real Data}\\
		%\toprule
		%\multicolumn{2}{c}{Test Data}   & \multicolumn{2}{c}{No Attack}  & 
		%\multicolumn{2}{c}{Grad. 1}&Grad. 2\\
		%\midrule
		% &Method  &Single& Ensemble  &Single& Ensemble  & Ensemble\\   \midrule
		\midrule
		\multicolumn{2}{c}{Model Specification}& Accuracy& \multicolumn{5}{c}{True Positive Rate / True Negative Rate}\\
		\midrule
		Method  & $h_{scale}$ & No Attack  & Plateau & Continuous  & Playback & Suppress & Flooding \\    \midrule
		CANet   &5  & 0.996 & \textbf{0.937} / 0.963  & \textbf{0.792} / 0.975   &  0.852 / 0.911  &  0.082 / 0.997 &  0.808 / 0.943 \\
		CANet  &10 & 0.994   &  0.913 / 0.988  & 0.701 / 0.985  &  0.878 / \textbf{0.977}  & 0.176 / 0.989  & 0.802 / 0.996  \\ 
		CANet  &20 &  0.995 & 0.936 / 0.968  & 0.724 / \textbf{0.988}  & 0.862 / 0.954 & \textbf{0.254} / 0.991 & 0.761 / 0.992 \\
		CANet  &30 & 0.992 & 0.936 / 0.970  & 0.740 / 0.983 & \textbf{0.903} / 0.940  & 0.240 / 0.983  & \textbf{0.833} / 0.994  \\
		\midrule 
		Predictive &- &0.995  & 0.269 / 0.949  & 0.577 / 0.985 & 0.134 / 0.964  & 0.001 / \textbf{0.998}  & 0.182 / \textbf{0.998}  \\
		Autoencoder  &- & \textbf{0.999} & 0.055 / \textbf{0.992}    &  0.491 / 0.983  &  0.079 / 0.977    &  0.007 / 0.995   &  0.627 / 0.996    \\
		\bottomrule \\
	\end{tabular}
	\caption{Summary of experimental results on real and synthetic CAN data. For normal test data the accuracy is recorded. For data with attacks the true positive rate, i.e. rate of attacks that are successfully detected, and the true negative rate, i.e. rate of normal data that is found as such, are recorded. Note that this is a pointwise evaluation criterion.} 
	\label{Eval_Table_real_Data}
\end{table*}

\begin{figure*}
	\begin{center}
		\includegraphics[scale=.25]{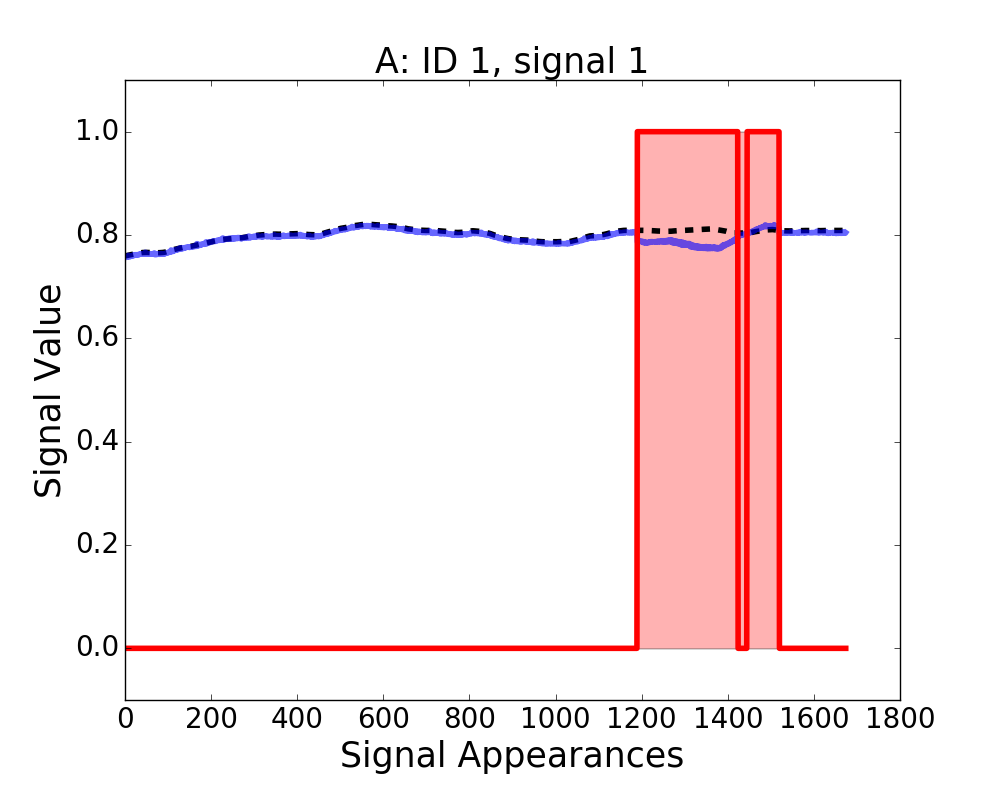}
		\includegraphics[scale=.25]{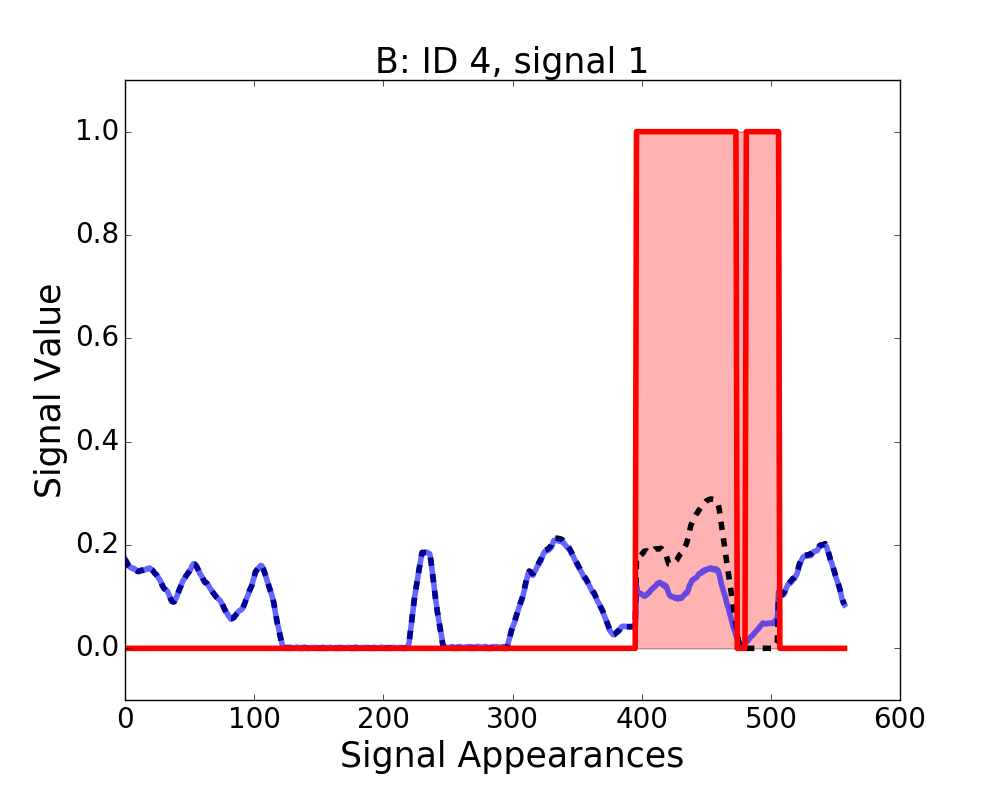}
		\includegraphics[scale=.25]{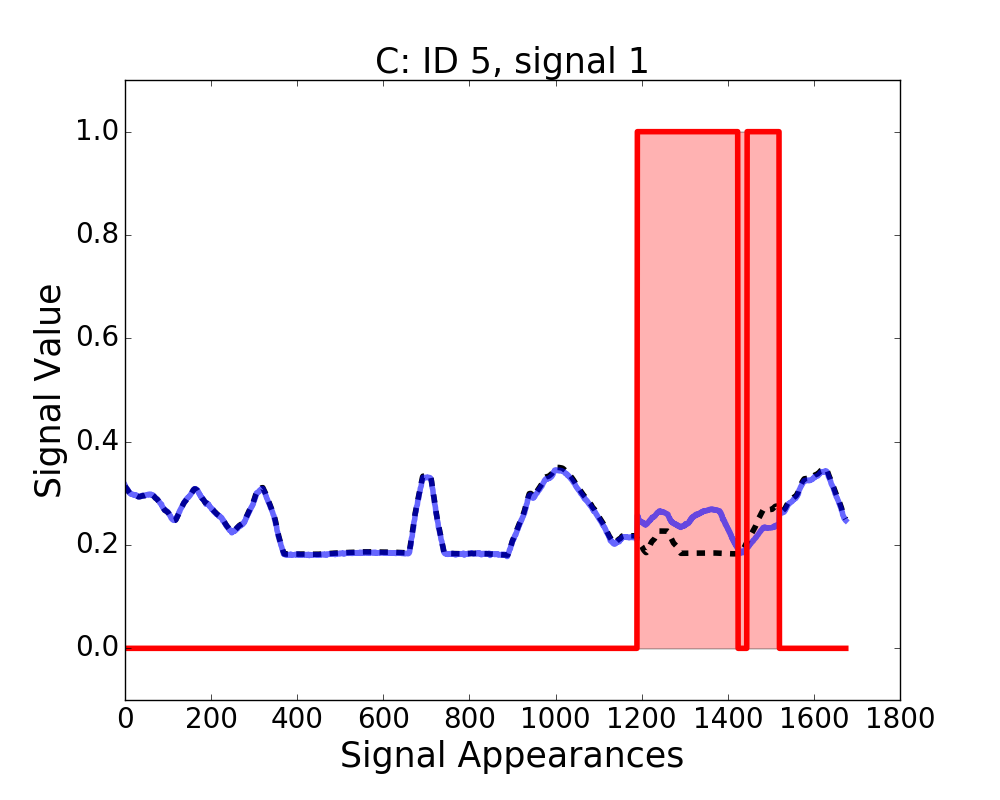}
		\includegraphics[scale=.25]{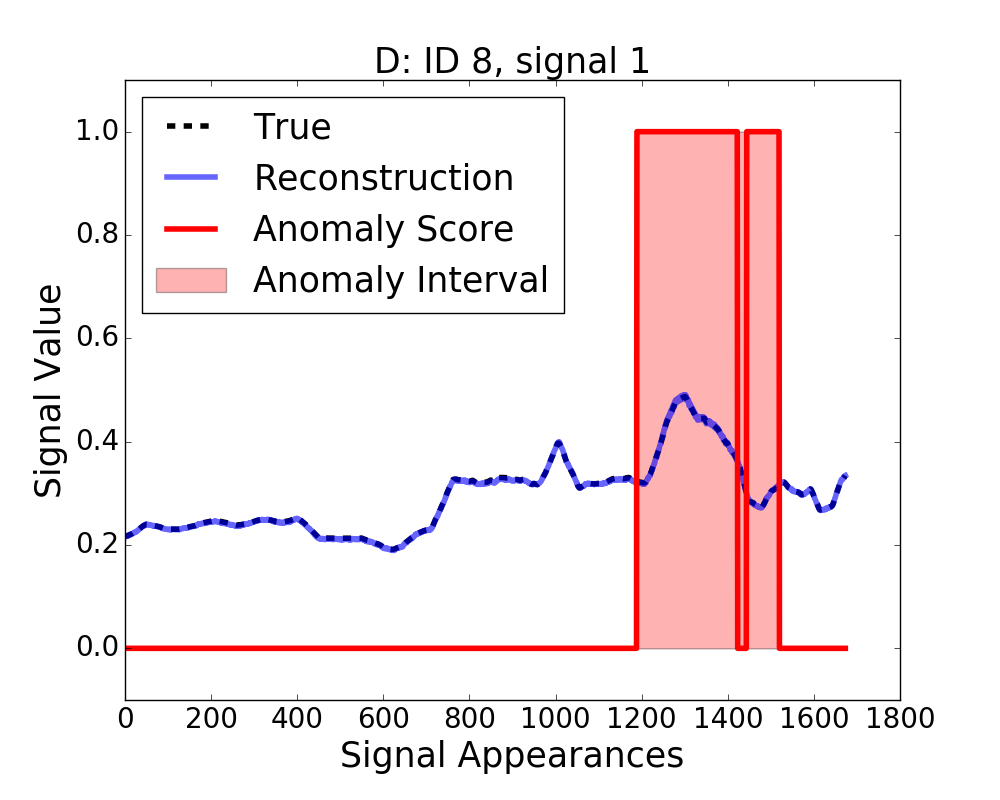}
	\end{center}

	\caption{The plots show four different signals of the synthetic CAN data set. They are extracted on the same time interval along with their corresponding reconstruction of CANet. The different number of signal appearances in $B$ compared with $A$, $C$ and $D$ are a result of different message frequencies. A playback attack on $B$ is performed. The corresponding attack interval is shaded in all plots. An intrusion is detected by CANet if the anomaly score (red line) equals one. It can be seen that on non-attacked data, the reconstruction (straight blue line) of the original signal (dashed black line) is almost exact. In the attack interval, deviations between the true signal and its reconstruction cannot only be observed in $B$ but also in $A$ and $C$. This is due to functional dependencies between these three signals. In contrast, $D$ has no functional dependencies with $A$, $B$ and $C$. Therefore, it is unaffected by the attack. Note that not the entire attack interval is detected. This is because the attacked signal is quite similar to the original signal for a short period of time. }
	\label{DataPics}
\end{figure*}

\begin{figure}
\begin{center}
\includegraphics[scale=.25]{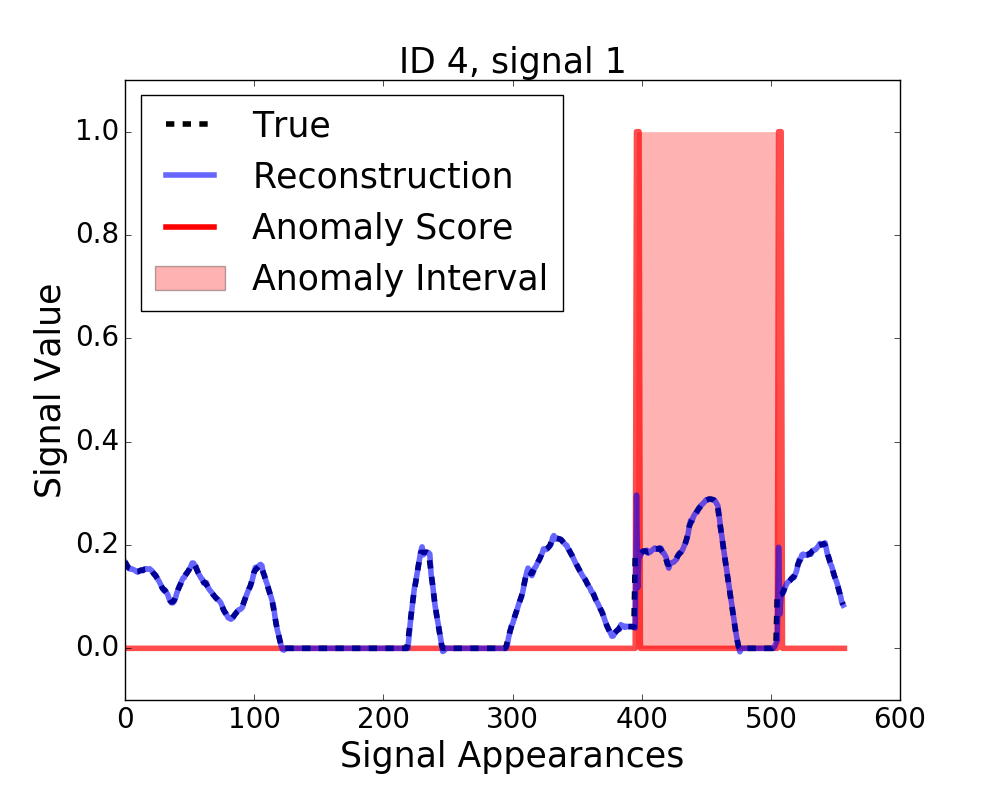}
\end{center}
\caption{Plot of the reconstruction as in Figure \ref{DataPics} using the predictive baseline. This method only detects the first few elements of the attacked interval correctly.}
\label{baslinePic}
\end{figure}

\begin{figure*}
	\begin{center}
		\includegraphics[scale=.25]{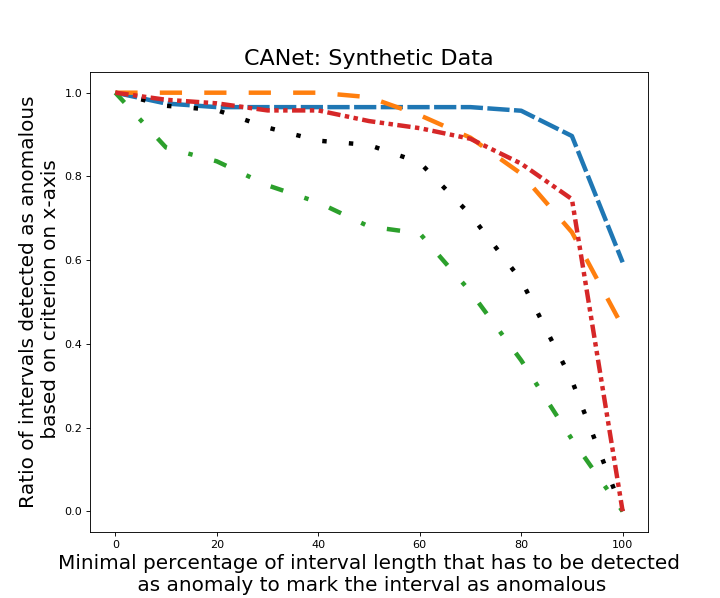}
		\includegraphics[scale=.25]{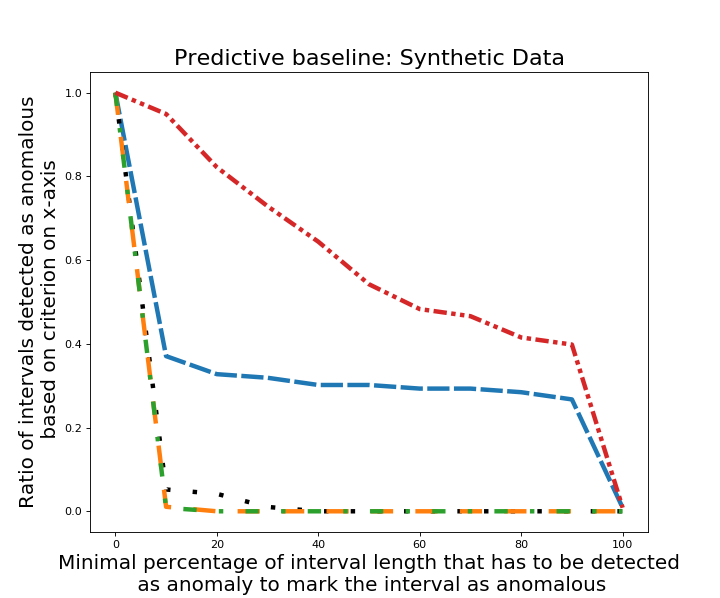}
		\includegraphics[scale=.25]{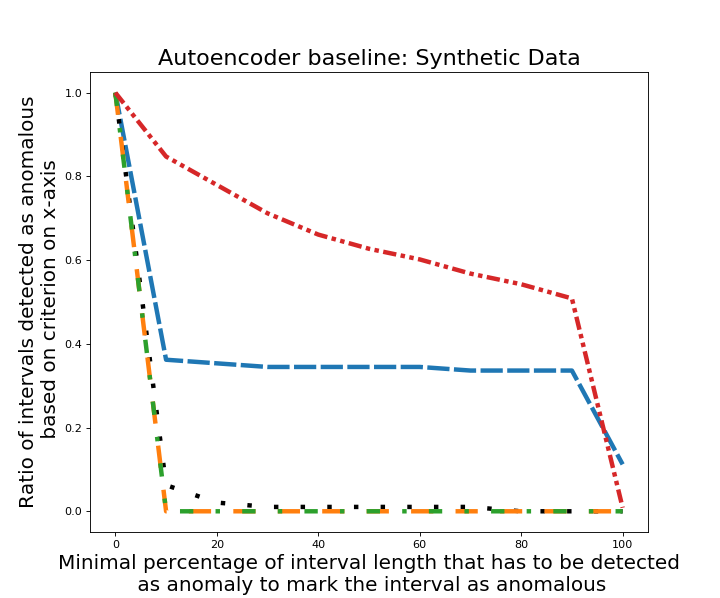}
		\includegraphics[scale=.25]{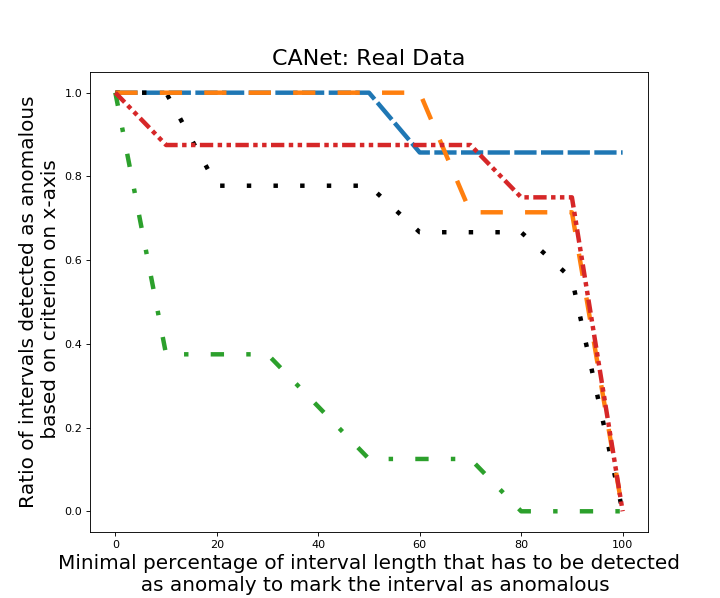}
		\includegraphics[scale=.25]{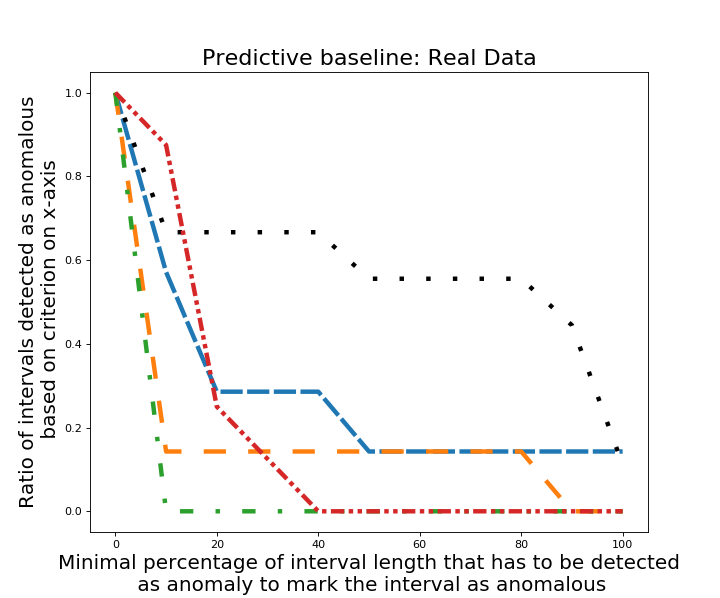}
		\includegraphics[scale=.25]{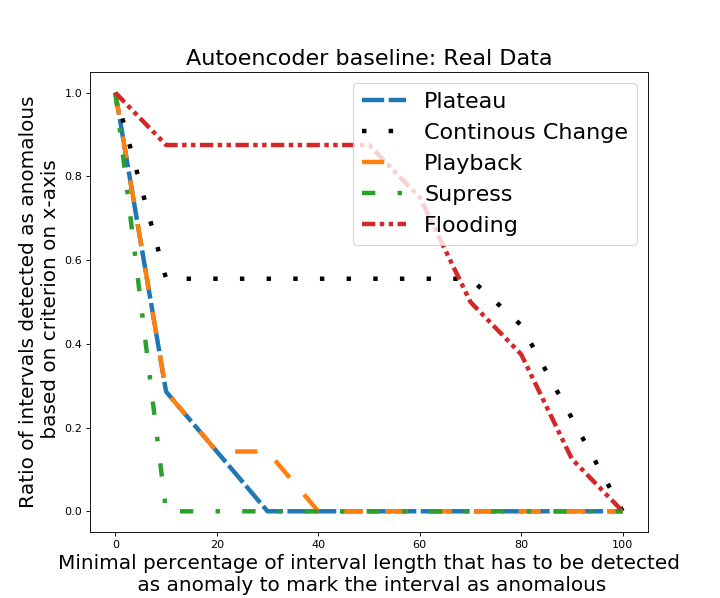}
	\end{center}
	\caption{The plot represents the ratio of detected attacks for each attack scenario. Here an attack is considered as the interval in which the attack is performed. The criterion for an attack interval to be detected is: At least $P\%$ of the interval have been point wise detected as attack (x-axis). The plots on the left hand side are based on our model with $h_{scale}=10$. By comparing the plots, it can be seen that CANet outperforms the other two approaches.}
	\label{Ratios}
\end{figure*}

\subsection{Evaluation}\label{eval_sec}

In this section, we describe  the numerical findings of CANet and compare them with the related research. For both, the real data and the synthetic data, we evaluate models of different $h_{scale}$ sizes  i.e. for $h_{scale} \in\{5,10,20,30\}$ (see Table \ref{net_Table}). Note that for comparability reasons no extra tuning of the network architecture for the different $h_{scale}$ values is made. A summary of the numerical results can be found in Table \ref{Eval_Table_real_Data}. The accuracies, true positive rates (i.e. rate of attacks that where successfully detected) and true negative rates (i.e. rate of normal data that was found to be normal) presented in the table are computed point wise. At the end of this section, we introduce an interval based evaluation criterion that might be more meaningful for real applications.

We find that in the real  as well as in the synthetic data setting the models identify normal data correctly in a solid way, with an accuracy typically larger than $0.99$. The attack types plateau, continuous change and playback  are detected reliably. The plateau and the playback attack show particular high detection rates, typically in the range of $[0.85, 0.955]$. The continuous change attack has a detection rate normally larger than $0.70$. Note that this is an excellent performance since the evaluation is done point wise. That is, we usually detect the vast majority of each attack interval (see Figure \ref{DataPics}). Our approach outperforms  the predictive and the autoencoder baseline by a significant margin. Typically, the baseline approaches only detect the first few attacked messages of the attack interval but identify the rest of the interval as normal (see Figure \ref{baslinePic}).  As expected, the baseline approaches perform particularly poorly on the playback attack. This is because if only a single signal of the CAN traffic is taken into account, deviations in the group of signals that have physical dependencies cannot be exploited. 

For comparison, we also evaluate our models on two other common attack types: suppress and flooding. These attacks can be detected by a rule based approach in a straight forward way, e.g. by analyzing the frequencies of each ID. Although our model does not have access to the time stamp and is therefore not specifically designed to find such attacks, it turns out that it still detects flooding attacks with a high true positive rate, whereas it struggles to detect suppress attacks. This is expected since the values that are added with a high frequency into the CAN bus during a flooding attack are much easier to be found than the gradual change of the network state that is the consequence of not sending a certain ID at all. When comparing with the baselines, we find that CANet is superior in all aspects. Nevertheless, both, the predictive and the autoencoder baseline, show relatively good results on the flooding attack. This is because in between the messages from the flooding attack the normal data points are still taken into account. Hence, during a single attack interval many anomalous large jumps in signal values can be found.

We find that the different choices for the parameter $h_{scale}$ have a relatively low effect on the performance of the models. Even small models with $h_{scale}=5$ perform reasonably well. This is especially interesting for a potential use of such models on an embedded device where memory and computational power are limited. Nevertheless, for the synthetic data set, we see a significant decrease in performance in the case $h_{scale}=30$. We believe this is due to overfitting. Of course, more parameters of the network architecture could be changed, e.g. the size of the autoencoder bottleneck, to prevent overfitting for large $h_{scale}$ values.

When comparing the models on the real and synthetic data, we find that the performance is in a similar range in most cases.  We believe that the synthetic data set is a good benchmark to test models for a CAN IDS, even if the data looks somewhat ``cleaner'' than in the real case. 

Some of the synthetic data is visualized in Figure \ref{DataPics}, which contains the plots of four different signals on the same time interval. In the signal of the upper right plot $B$, a playback attack has been injected. The attack interval is visualized by the shade in all plots. We observe that for all signals the reconstruction on normal data is usually really accurate, while during the attack interval deviations between the true data and its reconstruction can be found. Note that this deviation also appears in signals that are not explicitly attacked but only correlated in some way with the attacked signal, whereas signals without any correlations stay unaffected. Furthermore, over an attack interval typically not the entire attack is detected as such. This is expected, because an attacked signal and its original counterpart may have similar values in some parts of the attack window. For example, in case of a continuous change attack the modified signal values lie in a realistic range at the beginning of the attack. As a consequence, the model detects an attack only after the deviation between the original and the modified signal exceeds a certain threshold. 

Since we believe that in real application, finding a large number of attack intervals is more important than a high overall point wise accuracy on attacks (i.e. true positives), we investigate this by redefining what it means that an attack is found (see Figure \ref{Ratios}). That is, we define an attack as the entire period of time during which the attack is performed, i.e. an attack interval. We compute the percentage of attack intervals that are detected. Here, the criterion for identifying an interval as anomalous is that at least $P\%$ of that interval is detected point wise as anomaly. We can see in Figure  \ref{Ratios} that based on this definition CANet finds most anomaly intervals if $P\%$ does not get too large. This is true for both the real and the synthetic data case. However, both baseline methods have very low detection rates of anomaly intervals even for small $P\%$ (see Figure~\ref{Ratios}). 

Summing up the results, we find that CANet is capable of reliably detecting attacks on signals that have functional dependencies, while performing solidly on normal data. Our main findings are:
\begin{enumerate}
	\item The presented architecture is the first method that is capable of handling the difficult data structure from signals of multiple CAN IDs in a single model.
	\item CANet outperforms the selected baseline CAN IDS methods by a significant margin on all selected evaluation criteria.
	\item Our model has an excellent true negative rate and is capable of detecting unknown intrusions robustly. 
\end{enumerate}

\subsection{Risks and Benefits of Neural Network Based IDS Models}

The biggest advantage of using  machine learning based approaches, such as the one presented in this manuscript, is that they are potentially capable of detecting unknown intrusions. That is, they are successful in a task in which most other methods fail. Classically, for each possible attack scenario, a defense mechanism must be chosen. However, this process is highly time consuming and requires a significant amount of CAN bus domain expert knowledge for a successful detection. Here, neural networks significantly reduce both, the development time and the required CAN domain knowledge. 

On the other hand, the output of machine learning based methods can be complicated to analyze which makes it difficult to execute an automatic response once an intrusion is detected. Furthermore, neural networks require a large amount of trainig data and are typically more computational expensive than many other approaches. Another potential risk of using neural networks in a security setting is that they can be sensitive to adversarial attacks \cite{szegedy2013intriguing, tramer2017ensemble, strauss2017ensemble}.

\subsection{Reproducibility}

Typically, CAN bus based intrusion detection methods are tested on real data. However, publishing real CAN traffic and the corresponding CAN matrix is usually not possible, since it is considered intellectual property by most car producers. Hence, to the best of our knowledge, there is no standard data set for comparing methods. We try to close this gap by evaluating our model on both real and synthetic data and we make the synthetic data publicly available\footnote{The data is publicly available at \url{https://github.com/etas/SynCAN}.}. We hope that this simplifies the work of future researchers to compare their work with a baseline.

\section{Conclusion}\label{conc}
Cars are getting more and more connected. This opens ways for attacking the CAN bus of automobiles remotely. Since attacks can have a major impact on traffic safety, it is desirable that such attacks are detected in a robust manner. 

We present CANet, a novel neural network architecture that is trained in a unsupervised manner to detect intrusions and anomalies on the CAN bus. Furthermore, it is the first model in the literature capable of working on messages with different IDs simultaneously. The trained models have a high true negative rate, typically over $0.99$, which is necessary for real world applications. Furthermore, along with the high true negative rate we are able to detect a large amount of the unknown attacks, both on real and synthetic data. Our method is the only one in the literature capable of finding anomalies like the replay attack reliably. Although the results show the capabilities of the method, for applying it in real application further steps might be necessary. Those could include tuning the network architecture or redefining the anomaly score. 

For reproducibility of the method and in order to have a benchmark set for forthcoming approaches, our synthetic data is published at \url{https://github.com/etas/SynCAN}.

\begin{acks}

We thank Jens Gramm, Michael Oechsle and the ETAS Machine Learning group for the useful discussions. 

The views and conclusions contained in this document are those of the authors and should not be interpreted as representing the official policies, views or statements, either expressed or implied, of the affiliated organizations of the authors. 
\end{acks}

\bibliographystyle{ACM-Reference-Format}
%\bibliography{Literature}


\end{document}